\PassOptionsToPackage{dvipsnames}{xcolor}
\documentclass[manuscript]{acmart}
\usepackage{geometry}
\usepackage{pifont} % http://ctan.org/pkg/pifont
\usepackage[colorinlistoftodos]{todonotes}
\usepackage{amsmath}
\usepackage{blindtext}
\usepackage{xpatch}

\AtBeginDocument{%
  \providecommand\BibTeX{{%
    \normalfont B\kern-0.5em{\scshape i\kern-0.25em b}\kern-0.8em\TeX}}}

\setcopyright{acmcopyright}
\copyrightyear{2024}
\acmYear{2024}
\acmDOI{XXXXXXX.XXXXXXX}

\begin{document}

\title{Comparing the willingness to share for human-generated vs. AI-generated fake news}


%% TO DO: Correct the authors
\author{Amirsiavosh Bashardoust}
\email{amirsiavosh.bashardoust@gmail.com}
\orcid{0000-0001-9740-0905}
\affiliation{%
  \institution{University of Lausanne, Faculty of Business and
Economics}
  \streetaddress{Quartier de Chamberonne}
  \city{Lausanne}
  \country{Switzerland}
  \postcode{1015}
}
\author{Stefan Feuerriegel}
\email{feuerriegel@lmu.de}
\orcid{0000-0001-7856-8729}
\affiliation{%
  \institution{Munich Center for Machine Learning (MCML) \& LMU Munich}
  \streetaddress{} %TO DO
  \city{Munich}
  \country{Germany}
  \postcode{} %TO DO
}

\author{Yash Raj Shrestha}

\email{yashraj.shrestha@unil.ch}
\orcid{0000-0002-2699-4723}
\affiliation{%
  \institution{University of Lausanne, Faculty of Business and
Economics}
  \streetaddress{Quartier de Chamberonne}
  \city{Lausanne}
  \country{Switzerland}
  \postcode{1015}
}
\settopmatter{printacmref=false}
\setcopyright{none}
\renewcommand\footnotetextcopyrightpermission[1]{}
\pagestyle{plain}
\makeatletter
\let\@authorsaddresses\@empty
\xpatchcmd{\ps@firstpagestyle}{Manuscript submitted to ACM}{}{\typeout{First patch succeeded}}{\typeout{first patch failed}}
\xpatchcmd{\ps@standardpagestyle}{Manuscript submitted to ACM}{}{\typeout{Second patch succeeded}}{\typeout{Second patch failed}}    \@ACM@manuscriptfalse% Also in titlepage
\makeatother
\renewcommand{\shortauthors}{Bashardoust, Shrestha and Feuerriegel}

\begin{abstract}
Generative artificial intelligence (AI) presents large risks for society when it is used to create fake news. A crucial factor for fake news to go viral on social media is that users share such content. Here, we aim to shed light on the sharing behavior of users across human-generated vs. AI-generated fake news. Specifically, we study: (1)~What is the perceived veracity of human-generated fake news vs. AI-generated fake news? (2)~What is the user's willingness to share human-generated fake news vs. AI-generated fake news on social media? (3)~What socio-economic characteristics let users fall for AI-generated fake news? To this end, we conducted a pre-registered, online experiment with $N=$ 988 subjects and 20 fake news from the COVID-19 pandemic generated by GPT-4 vs. humans. Our findings show that AI-generated fake news is perceived as less accurate than human-generated fake news, but both tend to be shared equally. Further, several socio-economic factors explain who falls for AI-generated fake news.

\end{abstract}


\begin{CCSXML}
<ccs2012>
   <concept>
       <concept_id>10003120.10003121.10011748</concept_id>
       <concept_desc>Human-centered computing~Empirical studies in HCI</concept_desc>
       <concept_significance>500</concept_significance>
       </concept>
   <concept>
       <concept_id>10010405.10010455.10010461</concept_id>
       <concept_desc>Applied computing~Sociology</concept_desc>
       <concept_significance>300</concept_significance>
       </concept>
   <concept>
       <concept_id>10003120.10003130.10003131.10011761</concept_id>
       <concept_desc>Human-centered computing~Social media</concept_desc>
       <concept_significance>500</concept_significance>
       </concept>
 </ccs2012>
\end{CCSXML}

\ccsdesc[500]{Human-centered computing~Empirical studies in HCI}
\ccsdesc[300]{Applied computing~Sociology}
\ccsdesc[500]{Human-centered computing~Social media}

\keywords{fake news, misinformation, generative AI, large language model, online experiment, survey}

% TO DO: Check

\maketitle

\section{Introduction}

% AI can generate fake news

Generative artificial intelligence (AI) refers to technologies that can create human-like content \cite{feuerriegel_generative_2024}. As such, generative AI enables many applications for the better of society, such as simplifying information search \cite{dwivedi_opinion_2023}, supporting coding activities \cite{vaithilingam_expectation_2022}, and improving the accessibility for disabled people \cite{giri_exploring_2023}. However, generative AI can also be used for harmful applications where it produces fake news. Although the problem of fake news and deepfakes is not new \cite{toews_deepfakes_2023,van_der_nagel_verifying_2020,westerlund_emergence_2019}, recent advances in generative AI, such as GPT-4, have made it easy to produce fake news\footnote{We define fake news as ``fabricated information that mimics news media content in form but not in organizational process or intent`` \cite{lazer_science_2018}, thus implying that it can overlap with other, related terms such as misinformation and disinformation.} that is highly realistic and thus hard to detect by humans \cite{spitale_ai_2023}. To this end, AI-generated fake news presents immediate threats to society.

% characteristics of AI-generated fake news

The threat vectors of AI-generated fake news are characterized by three main factors  \cite{feuerriegel_research_2023}, namely, scale, speed, and usability, as follows. (1)~AI can be used to automate the mass production of fake news, likely leading to an uncontrollable influx of fake news which is difficult to detect and moderate. For example, AI tools can simply generate dozens of different fake news with different content around the same story. (2)~AI can create fake news in seconds and can thus outpace traditional fact-checking. This is different from human-generated fake news, which requires time for writing and editing. (3)~AI tools have become widely available, reducing the entry barriers for a broad user base to generate fake news without the need for specialized skills. As a consequence, it is crucial to prepare for the potential risks in an upcoming era of AI-generated fake news by understanding how users will respond to AI-generated fake news, so that effective, evidence-based mitigation strategies can be developed.

Previous works have studied the characteristics of AI-generated content and what sets it aside from human-generated content. For example, one work compares human-generated vs. AI-generated content in terms of linguistic characteristics (e.g., use of slang, use of emotions) and semantic characteristics (e.g., level of detail, communication of uncertainties) \cite{zhou_synthetic_2023}. Others have analyzed the heuristics with which humans discern human-generated vs. AI-generated content \cite{spitale_ai_2023,jakesch_human_2023,kreps_all_2022}, finding that it is generally difficult for humans to infer where content is coming from \cite{jakesch_human_2023}. Furthermore, Spitale et al. \cite{spitale_ai_2023} evaluated whether users can predict whether true or false content was generated by humans or by an AI, finding that AI-generated content can disinform more effectively than compared to human-generated content. Hence, while the perception of false information was studied before, research studying the behavior of users interacting with false information is lacking. In particular, it is unclear whether users have a different willingness to share for human-generated vs. AI-generated fake news. Importantly, sharing activities for fake news are not necessarily related to the perception of fake news (e.g., even though some users disbelieve the content of fake news, they may share it nevertheless due to a lack of attention) \cite{pennycook_fighting_2020}. 

% motivation 

There are good reasons why to expect that users share human-generated vs. AI-generated fake news differently. (a)~\emph{Human-generated fake news is shared more.} Generating fake news by AI requires not only mimicking human-like writing but also ideation of incorrect facts in the first place, which may be not sufficiently convincing. Hence, users do not believe the AI-generated fake news and thus do not share it. (b)~\emph{AI-generated fake news is shared more.} Generative AI writes text in a highly structured and error-free manner, which makes it more convincing and thus increases the propensity of users to believe in AI-generated fake news and share it. (c)~\emph{Human-generated and AI-generated fake news are shared similarly}. Since users cannot discern whether texts were generated by AI or humans, it is likely that online users will fall for both human-generated and AI-generated fake news and thus share both at equal rates. Motivated by this, we study the following three \textbf{research questions (RQs)}:
\begin{itemize}
\item \textbf{RQ1:} \emph{What is the perceived veracity of human-generated vs. AI-generated fake news?}
\item \textbf{RQ2:} \emph{What is the willingness to share of human-generated fake news vs. AI-generated fake news?}
\item \textbf{RQ3:} \emph{Which socio-economic characteristics are determinants that explain which users fall for AI-generated fake news?}
\end{itemize}

% what we do

\noindent
In this paper, we conducted a pre-registered, online experiment ($N=988$ subjects) to compare how users respond to human-generated vs. AI-generated fake news.\footnote{Code and data supporting our findings are available at \href{https://github.com/vosh-96/Comparing-the-willingness-to-share-for-human-generated-vs.-AI-generated-fake-news/tree/main}{public Git repository}. The pre-registration protocol is available at \url{https://osf.io/vhnju?view_only=604564b14cf2471789b099a3164d109f}} In our experiment, we analyze two important variables characterizing online behavior around fake news, namely, perceived veracity and willingness to share. While perceived veracity has been studied earlier \cite{spitale_ai_2023}, we add by understanding the sharing behavior. Our experiment is based on a corpus of 20 fake news generated by GPT-4 and 20 by humans. We chose fake news from the COVID-19 pandemic due to the fact that wrong beliefs about it are still widespread and such wrong beliefs had large, negative effects for public health. We further perform regression analyses to identify various socio-economic determinants that characterize users likely to fall for fake news generated by AI. We find that AI-generated fake news is perceived as less accurate than human-generated fake news, but both tend to be shared equally.

%Background

\section{Related Work}

\subsection{Fake News on Social Media}

The phenomenon of fake news poses a great threat to the integrity of online information ecosystems and, more broadly, the functioning of modern societies. Among others, fake news has been used to meddle with elections \citep{allcott_social_2017}, manipulate stock markets \cite{clarke_fake_2020}, and hamper public health \cite{shirish_impact_2021, schuetz_combating_2021}. Furthermore, fake news circulated widely on social media during the COVID-19 pandemic and, as a result, caused several thousand deaths that could have been prevented \cite{shirish_impact_2021, schuetz_combating_2021}. Still, the effective detection and mitigation of fake news present ongoing challenges  \cite{zhou_survey_2021}.

% STEFAN: we need a bit more here

Extant research has examined the relationship between perceived veracity and subsequent sharing of traditional, human-generated fake news (e.g., \cite{pennycook_psychology_2021}). Different elements in social media could distract the attention of users from judging the perceived veracity, which will affect their decision to share \cite{pennycook_psychology_2021}. This has important implications for our paper: simply studying the perceived veracity of human-generated vs. AI-generated fake news does \underline{not} inform how human-generated vs. AI-generated fake news is shared online. The latter is further relevant given that sharing is crucial for fake news (or any social media content more broadly) to go viral and eventually reach a large audience \cite{maarouf_virality_2023,geissler_russian_2023}.

\subsection{Generative AI for Content Creation}  

Recent advances in the capabilities of generative AI make it easy to produce content almost indistinguishable from that of humans \cite{jakesch_human_2023,schuetz_combating_2021}. For instance, the Generative Pre-trained Transformer 3 (GPT-3) is able to produce a wide range of texts including writing lyrics for songs, completing documents, programming code, fictional stories, and cooking recipes \cite{kobis_artificial_2020}. As another example, large language models have reached a level of performance that ad content created by generative AI is perceived to be of higher quality compared to ad content produced by human experts \cite{zhang_human_2023}.

Research has also analyzed how users respond to AI-generated content \cite{jakesch_human_2023}, but outside of fake news. For example, research shows that people cannot distinguish between human-generated and AI-generated news stories \cite{kreps_all_2022}. Moreover, people who are more confident in detecting AI content made more errors in distinguishing both \cite{miller_ai_2023}. In another study, users rated news headlines as less accurate when they were told that the same headline was written by AI rather than by humans \cite{longoni_news_2022}. Others found that AI-generated content can be more persuasive than human-generated content, for example, in the context of propaganda \cite{goldstein_can_2023} and letters to politicians \cite{kreps_potential_2023}. Here, the underlying rationale is that AI lacks human sociological and psychological elements, such as individual desires and emotions, because of which outputs may be perceived to be less personal but more accurate. These features can also increase the perceived accuracy of AI-generated output more broadly \cite{gray_dimensions_2007,gray_feeling_2012,jago_assumptions_2022}. However, the previous works focus on regular content (e.g., letters, dating profiles) and \underline{not} fake news.

Despite the potential harm of AI-generated fake news, there is, so far, little research exploring how users respond to AI-generated fake news, especially in comparison to human-generated fake news. One work has identified the linguistic differences between human-generated and AI-generated fake news \cite{zhou_synthetic_2023}. However, the aforementioned analysis is focused on significant linguistic differences in expression patterns between human-generated and AI-generated fake news but \emph{not} how users perceive the veracity of each. Even others focus on whether users can predict the source of fake news \cite{spitale_ai_2023}. However, research aimed at understanding the users' willingness to share fake news has remained unclear, even though is an essential precondition for fake news to go viral. 
 
\vspace{0.2cm}
\textbf{Research gap:} The perception of AI-generated content by users has been widely studied, but still, salient questions about how users engage with AI-generated fake news are open (Table~\ref{tab:gap}). To this end, we provide the first experimental study comparing the willingness to share of users across human-generated and AI-generated fake news. 

\begin{table} 
\caption{Key literature and how our study is different.} 
\label{tab:gap} 
\renewcommand{\arraystretch}{1.3}
\footnotesize
\begin{tabular}{lllcc} 
\toprule
Reference & Scope & Content &\multicolumn{2}{c}{Outcomes} \\
\cmidrule(lr){4-5}
 &&& Perceived veracity & Willingness to share\\ 
\midrule 
Pennycook et al. \cite{pennycook_fighting_2020} & Human-generated  & True vs. fake news & \textcolor{ForestGreen}{\ding{52}} & \textcolor{ForestGreen}{\ding{52}} \\
Spitale et al. \cite{spitale_ai_2023} &  Human-generated vs. AI-generated & True vs. false Twitter/X posts & \textcolor{ForestGreen}{\ding{52}} & \textcolor{BrickRed}{\ding{54}} \\
\midrule
\textbf{Our study} & Human-generated vs. AI-generated & Fake news & \textcolor{ForestGreen}{\ding{52}} & \textcolor{ForestGreen}{\ding{52}} \\
\bottomrule
\end{tabular}
\end{table}

\section{Method}

The experimental protocol was pre-registered at \href{https://rb.gy/qu8bu}{https://rb.gy/qu8bu}. Ethics approval was granted by the Institutional Review Board at the University of (name anonymized for review).

\subsection{Experimental Task}

% stimulus

We conducted a within-subject online experiment, where subjects ($N = 988$) evaluated 20 items with fake news related to the COVID-19 pandemic. Out of the 20 items, 10 were human-generated and 10 were AI-generated fake news (details are in Sec.~\ref{sec:materials}). The order in which the fake news items were displayed was randomized. Furthermore, subjects were neither informed that the content they were going to read was fake news nor that content was generated by AI.

% DVs

Subjects were asked to respond to two questions for each fake news item: 
\begin{enumerate}
\item \emph{Perceived veracity:} Subjects were asked to rate the perceived veracity of each item based on a 5-point Likert scale from 0 (\emph{``strongly inaccurate''}) to 4 (\emph{``strongly accurate''}), where 2 denotes a neutral assessment. 
\item \emph{Willingness to share:} Subjects were asked to report their willingness to share the news item on their social media. They could choose between \emph{``not sharing''} (encoded as 0), \emph{``sharing''} (encoded as 1), and \emph{``don't know''} (excluded from analysis). 
\end{enumerate}
At this point, we highlight that the perceived veracity and the willingness to share measure different concepts. Importantly, the correlation coefficient between both is 0.162, which is referred to as weak correlation.

\subsection{Subjects}

% recruiting

Subjects were recruited from the U.S. via the online platform \emph{Lucid} (\url{https://luc.id/}). A particular benefit of the Lucid platform is that it maintains a sample of subjects from the U.S. that is representative in terms of different demographic factors such as race, age, and gender. In our experiment, we intentionally focused on subjects from the U.S. to ensure that all subjects have a basic literacy in the English language. We did not compensate subjects directly; instead, subjects received a fixed compensation through \emph{Lucid} for successful completion. The overall cost was EUR~1536. Given that the survey took only around 10 minutes to complete, the compensation of subjects is in alignment with the minimum federal wage \cite{noauthor_minimum_2023}.

% sample size

We aimed for $N=$ 1000 subjects. The choice was informed by earlier studies that analyze the perceived veracity and other determinants for human-generated fake news \cite{pennycook_understanding_2020,pennycook_fighting_2020}. Accordingly, 3610 subjects were initially recruited for our survey, out of which only 1185 of them passed the attention tests (described later) and completed the survey. Out of them, 193 subjects declared during the survey that they answered at least one question randomly or searched the Internet for the shown fake news, because of which we excluded the 193 subjects. We further excluded four subjects because they entered to be below the minimum age of 18 years, which was enforced by our Institutional Review Board. Eventually, this led to 988 remaining subjects, which is close to our planned $N=$ 1000 subjects.

% descriptives

Out of the 988 subjects, 626 were female, 342 were male, and 20 selected other genders. In our sample, 78.92\% identified themselves as white, and 11.25\% as Black or African American. Our sample has a mean age of 47.53 years (SD $=$ 17.13). Out of our 988 subjects, 74.5\% have been vaccinated against COVID-19. When considering political orientation, 19.23\% identified as left, 48.69\% as moderate, 27.63\% as right, and 4.45\% as other. In terms of social orientation, 32.7\% identify as liberal, 38.6\% as moderate, and the remaining 28.7\% as conservative. Details about the questions are in Table~\ref{tab:detailQues}.

\subsection{Materials}
\label{sec:materials}

We created a corpus of 10 human-generated and 10 AI-generated fake news around the COVID-19 pandemic. We chose the COVID-19 pandemic for our study due to the fact there there is still a widespread belief in COVID-19 fake news among society, with severe negative impact on public health \cite{hsu_as_2022}. Our process was as follows:
\begin{itemize}
\item \emph{Human-generated fake news.} We followed previous research (e.g., \cite{vosoughi_spread_2018,pennycook_fighting_2020,lazer_science_2018,allcott_social_2017,lutz_affective_2023,lutz_which_2024} and retrieved fake news from independent fact-checking websites, namely, {PolitiFact} (\url{https://www.politifact.com}) and {Snopes} (\url{https://www.snopes.com}). PolitiFact and Snopes are fact-checking websites that rate the accuracy of claims made by elected officials, candidates, political parties, and other groups. Their fact-checking process involves researching statements, interviewing experts, and then providing a detailed analysis of the claim. For our corpus, we collected items with a final rating of \emph{``pants on fire''} or \emph{``false''} at PolitiFact and \emph{``false''} at Snopes. Besides these two sources, we also used COVID-19 fake news from a study by Pennycook et al.~\cite{pennycook_fighting_2020}. Overall, we used four fake news from Snopes, two from PolitiFact, and the rest from  Pennycook et al.~\cite{pennycook_fighting_2020} research. The rationale behind our sampling strategy spanning different sources is to generate a more representative sample that also captures sufficient diversity among fake news items that proliferate online.\footnote{We performed robustness checks where we repeated the analysis for fake news from PolitiFact only or for fake news from Snopes only, yielding consistent findings.}
\item \emph{AI-generated fake news.} The sample of AI-generated fake news was created using GPT-4.\footnote{We used the version from 14-03-2023, which we accessed via \emph{Bing Chat}.} Specifically, we asked GPT-4 to create fake news around the COVID-19 pandemic and to use a format similar to social media. Our corresponding prompt was as follows: \emph{``Write an incorrect statement about COVID-19 and put it in quotation marks.''} At the time when we accessed GPT-4, there were not safety measures that blocked our prompts. As a result, we were able to simply generate fabricated content with misleading information as outputs to our prompts. Currently, stricter safety measures are in place, and, hence, researcher would need to use alternative strategies to convince GPT-4 to output fabricated information, such as through impersonation techniques \cite{spitale_ai_2023}. 

\end{itemize}
Table~\ref{tab:example_fakenews} shows examples of human-generated and AI-generated fake news. The complete list is in the Online Supplements. Importantly, the human-generated vs. AI-generated fake news have a similar length to ensure comparable reading effort.

\begin{table} 
\caption{Examples of fake news in our corpus. Full list in the Online Supplements.} 
\label{tab:example_fakenews} 
\renewcommand{\arraystretch}{1.3}
\footnotesize
\begin{tabular}{ll} 
\toprule
Content & Label \\ 
\midrule 
\emph{``Federal Emergency Management Agency proposes martial law to contain virus.''} & Human-generated \\
\emph{``Coconut oil's history in destroying viruses, including coronaviruses.''} & Human-generated \\
\emph{``The second booster has 8 strains of HIV.''} & Human-generated \\
\midrule
\emph{``The coronavirus only spreads in cold weather or places with low humidity.''} & AI-generated \\
\emph{``COVID-19 vaccine causes infertility in women and men.''} & AI-generated \\
\emph{``COVID-19 is less deadly than the flu and does not require any special measures.''} & AI-generated \\
\bottomrule
\end{tabular}
\end{table}

\begin{figure}[htbp]
  \centering
  \includegraphics[width=0.8\linewidth]{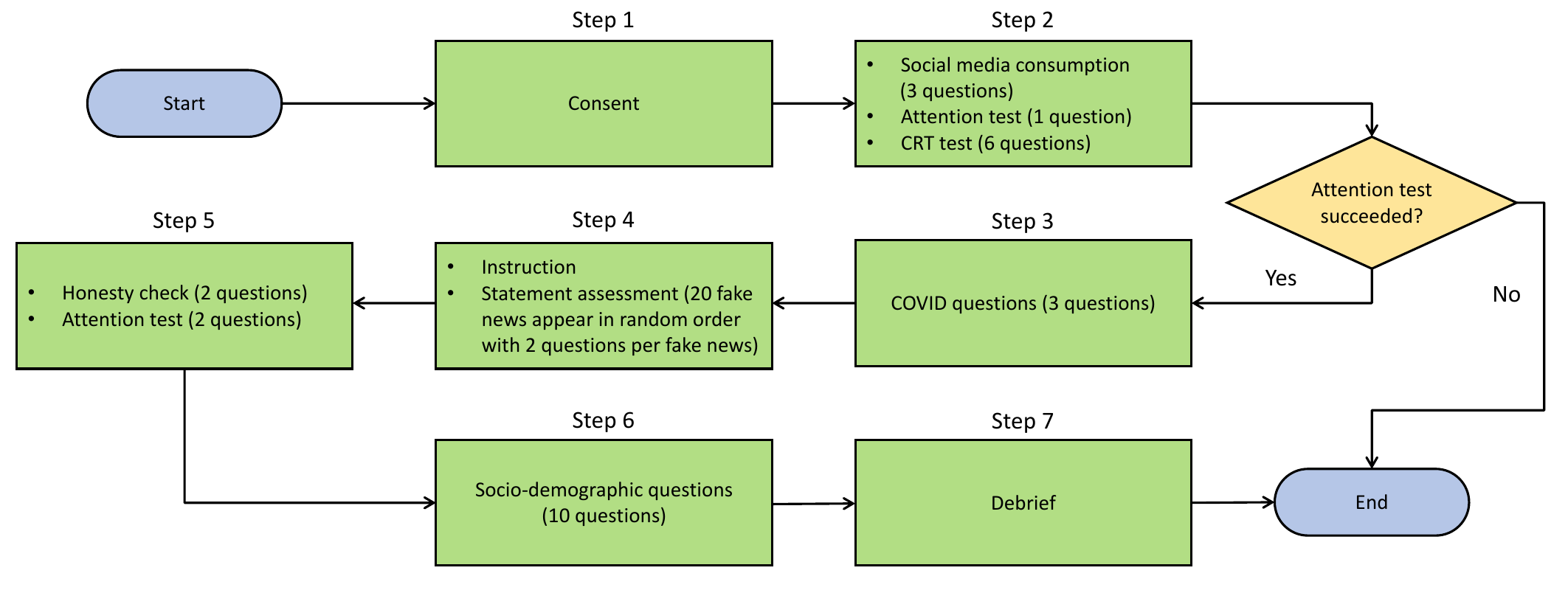}
  \caption{The experiment's procedure flowchart.}
  \Description{Flow chart showing the procedure of our experiment.}
  \label{fig:flowchart}
\end{figure}

\subsection{Procedure}

% STEFAN: table with questions

Our experimental procedure consists of seven steps. Figure \ref{fig:flowchart} shows the flowchart of our experimental task. 

\emph{Step 1.} Before the start of the experiment, subjects first visited a landing page. Subjects had to give informed consent in order to proceed.

\emph{Step 2.} Subjects were asked about their social media consumption. The first question is what kind of content they share on social media (i.e., users could select multiple options out of \emph{``Political news''}, \emph{``Sport news''}, \emph{``Celebrity news''}, \emph{``Science/technology news''}, \emph{``Business news''} and \emph{``Others''}). Then, we asked them what social media platform they use (i.e., users could select multiple options out of \emph{``Facebook''}, \emph{``Twitter''}, \emph{``Snapchat''}, \emph{``Instagram''}, \emph{``WhatsApp''}, \emph{``I do not use social media''}, and \emph{``Others''}). Finally, we asked them how much time (in hours) they spend, on average, on social media per day.

Furthermore, we asked subjects to attend an attention test (detailed in Table \ref{tab:detailedAttentionTest}) in which we asked subjects to report which news outlet they regularly check, yet the question concluded by stating that the question should be ignored and a specific option should be chosen. If the attention test was failed, the survey was terminated immediately, and the subject was excluded from our analysis to ensure that subjects read all texts thoroughly, which is analogous to other online experiments \cite{pennycook_fighting_2020}. 

Step 2 also included a six-item cognitive reflection test (CRT) \cite{frederick_cognitive_2005}. Following  Penneycook et al.~\cite{pennycook_fighting_2020}, we used the modified version of the original three CRT questions. This modification includes three additional non-numeric items and has reworded questions. The CRT evaluates an individual's tendency to be intuitive \cite{pennycook_is_2016,toplak_cognitive_2011}. Each CRT question is designed to trigger an intuitive yet incorrect answer. For instance, one of our CRT questions is: \emph{``A farmer had 15 sheep, and all but 8 died. How many are left?''} The intuitive answer one might think of is 7, but the correct answer is 8. The ability to think critically is associated with answering CRT problems correctly. The items are aggregated, so that CRT scores range between 0 (if all answers are incorrect) to 6 (if all responses are correct).

\emph{Step 3}. We asked subjects three questions regarding the COVID-19 pandemic. The first question was how concerned they were about the virus during the pandemic (as a percentage from 0\% to 100\%). The second question was how often they checked the news about COVID-19 during the pandemic on a 5-point Likert scale from 1 (\emph{``never''}) to 5 (\emph{``very often''}). Finally, we asked them whether they had been vaccinated against COVID-19 (\emph{``yes''} or \emph{``no''}). Altogether, this was done to later assess their prior disposition toward information around the COVID-19 pandemic \cite{kreps_all_2022,pennycook_fighting_2020}.

% to assess their prior disposition .

\emph{Step 4.} We first provided subjects with instructions about the experimental task, informing that they will face a series of statements about COVID-19 and they should answer two questions for each statement. We then showed them 20 different fake news items that were either human-generated or AI-generated  in random order. For each, we asked users to assess (1)~the accuracy of each item and (2)~the willingness to share the item. We again emphasize that the subjects were not aware that all items were fake news. During Step~4, we also monitored the time that users spent on each item to later analyze it as a proxy of attention and cognitive involvement. 

We performed two more attention tests after the end of this step to check if subjects still are actively involved in the experiment. However, unlike the first attention test, we did not terminate the experiment and they were able to continue but excluded them from our analysis.\footnote{We further ran a robustness check, where the subjects were included, but arrived at conclusive findings.}

\emph{Step 5.} After completion of the experimental task, we followed previous research \cite{pennycook_prior_2018} and asked subjects if, at any point, they answered a question randomly or if they searched the Internet (``honesty tests''). We excluded those subjects who answered this question with a "yes".\footnote{We repeated our analyses with the inclusion of these subjects and arrived at consistent conclusions.} We also conducted another attention test, which consisted of two questions: (1)~\emph{``5 + 4 = 11''}; and (2)~\emph{``The year 1820 came before the year 1910''}. For both questions, subjects had to rate the items using 5-point Likert scale from 1 (\emph{``strongly disagree''}) to 5 (\emph{``strongly agree''}). We excluded subjects who failed any of these two questions.

\emph{Step 6.} We further asked various questions about socio-demographics such as age, gender, years of education, ethnicity, social and economic orientation (from 0 \emph{``strongly liberal''} to 4 \emph{``strongly conservation''}), religiosity (\emph{``religious person''}, \emph{``not a religious person''}, and \emph{``atheist''}), political orientation (\emph{``left''}, \emph{``moderate''}, \emph{``right''}, and \emph{``other''}), fluency in English, and annual income.

\emph{Step 7}. We debriefed the subjects. In particular, all subjects were informed that all statement that they had seen during the experiment were fake and that subjects should not rely on any of them.

The detailed questions and their possible answers are listed in Table \ref{tab:detailQues}.
\begin{table}[h!] 
\caption{Detailed survey questions.} 
\footnotesize
\label{tab:detailQues} 
\begin{tabular}{lll  l} 
\toprule
  \textbf{Variable name} & \textbf{Question / measurement procedure} &  \textbf{Options}  \\ 
\midrule 
 \textsc{Dependent variables} & \\ 
 \midrule
 \emph{PerceivedVeracity} & \emph{``How do you rate the accuracy of the statement below?''} &  very inaccurate (0) / ... / very accurate (4) \\ 
 \emph{WillingnessToShare} & \emph{``Will you share this statement on your social media''}? & yes (1) / no (0) / don't know ($-$1) \\
\midrule
 \textsc{Independent variables} & \\
 \midrule
 \emph{VaccinationStatus} & \emph{``Have you been vaccinated against COVID-19?''} & yes (1) / no (0)\\
 \emph{Concern} & \multicolumn{1}{p{5.6cm}}{\emph{``How concerned were you about COVID-19 pandemic?''} } & [0\% (not concerned), 100\% (extremely concerned)]\\  
 \emph{NewsConsumption} & \multicolumn{1}{p{5.6cm}}{\emph{``How often did you proactively check COVID-19 pandemic news?''}} & never (0) / 1 / ... / very often (4) \\
 \emph{CRT} & \multicolumn{1}{p{5.6cm}}{ \emph{``The ages of Mark and Adam add up to 28 years total. Mark is 20 years older than Adam. How many years old is Adam?''} \newline
 \emph{``If it takes 10 seconds for 10 printers to print out 10 pages of paper, how many seconds will it take 50 printers to print out 50 pages?''} \newline
 \emph{``On a loaf of bread, there is a patch of mold. Every day, the patch doubles in size. If it takes 40 days for the patch to cover the entire loaf of bread, how many days would it take for the patch to cover half of the loaf of bread?''} \newline
 \emph{``If you’re running a race and you pass the person in second place, what place are you in?''} \newline
 \emph{``A farmer had 15 sheep and all but 8 died. How many are left?''} \newline
 \emph{``Emily’s father has three daughters. The first two are named April and May. What is the third daughter’s name?''} } & all incorrect (0) / ... / all correct (6) \\
 \emph{Age} & \emph{``What is your age? (in years)''} & - \\ 
 \emph{Gender}$^{*}$ & \emph{``What is your gender?''} & \multicolumn{1}{p{5.2cm}}{ male (0) / female (1) / transgender female / transgender male / trans or non binary / not listed / prefer not to say} \\
 \emph{Fluency} & \emph{``Are you fluent in English?''} & yes (1) / no (0)\\
 \emph{PoliticalOrientation} & \multicolumn{1}{p{5.6cm}}{\emph{``Which of the following best describes your political position?''}} & left (0) / moderate (1) / right (2) / other (3) \\
 \emph{SocialOrientation} & \emph{``On social issues I am:''} & strongly liberal (0) / 1 / ... / strongly conservative (4) \\ 
 \emph{EconomicOrientation} & \emph{``On economic issues I am:''} & strongly liberal (0) / 1 / ... / strongly conservative (4) \\ 
 \emph{EducationYears} & \multicolumn{1}{p{5.6cm}}{\emph{``How many years of formal education have you completed?''}} & 0 / ... / 20] \\
\emph{Religiosity} & \emph{``Would you say you are:''} & religious person (0) / not a religious person (1) /\\ 
&& an atheist (2) \\
\emph{Income}$^{\dagger}$ & \emph{``What is your household income (before taxes)?''} & \multicolumn{1}{p{5.2cm}}{ [\$10,000, \$19,999] / \ldots / [\$150,000 or more] \newline
[\$10,000, \$29,999]: low (0) / [\$30,000, \$79,999]: middle (1) / $\geq$ \$80,000: high (2)}\\
\emph{Ethnicity}$^{**}$  & \emph{``Which category best describes you?''} 
& \multicolumn{1}{p{5.2cm}}{ White / Hispanic, Latino or Spanish origin / Black or African American / Asian / American Indian or Alaska Native / Middle Eastern or North African / Native Hawaiian or Other Pacific Islander / some other race, ethnicity or origin; encoded as White (0) / Non-White (1)}\\
\midrule

\multicolumn{3}{p{14.1cm}}{$^*$ Only 20 subjects identified themselves as being different from male or female. Therefore, in our regression, we encoded \emph{Gender} as a binary variable and remove other observations from the analysis due to the low statistical power. We later allow for between-subject variation beyond binary variables through random effects. }\\
\multicolumn{3}{p{14.1cm}}{$^\dagger$  We transform our income values into three categories of incomes: low, middle, and high. This coding is based on \cite{bennett_improving_2021}; using \$80,000 as the threshold between middle and high. }\\
\multicolumn{3}{p{14.1cm}}{$^{**}$ In our main analysis, we encoded \emph{Ethnicity} as a binary variable due to the low number of different identified ethnicities. }

\end{tabular}
\end{table}

\begin{table}[h!] 
\caption{List of attention and honestly tests.} 
\footnotesize
\label{tab:detailedAttentionTest} 
\begin{tabular}{l  l} 
\toprule
  \textbf{Questions} &  \textbf{Options}  \\ 
\midrule
 \textsc{Attention checks} & \\ 
  \midrule
  \emph{``When a big news story breaks, people often go online} & New York Times / Yahoo! News /\\ \emph{to get up-to-the-minute details on what is going on. We also want} & Huffington Post / NBC.com / CNN.com /\\  
  \emph{know which websites people trust to get this information.} & USA Today / FoxNews.com / Others \\  
  \emph{ to know if people are paying attention to following questions.} \\ 
  \emph{Please ignore the question and select FoxNews.com} \\
  \emph{and NBC.com as your two answers.''} \\   
 \emph{`` 5 + 4 = 11''} & strongly disagree (0) / 1 / ... / strongly agree (4) \\  
 \emph{``The year 1820 came before the year 1910.''} & strongly disagree (0) / 1 / ... / strongly agree (4)\\

\midrule
  \textsc{Honesty tests} & \\
   \midrule
   \emph{``Did you respond randomly at any point during the study? ''} &  yes / no\\
  \emph{``Did you search the internet (via Google or otherwise) for help?
''} & yes / no\\
\bottomrule
\end{tabular}
\end{table}

\subsection{Statistical Analysis}

To answer our research questions, we conducted the following statistical analyses:
\begin{enumerate}
\item For RQ1, we performed hypothesis testing to compare the \emph{PerceivedVeracity} across human-generated vs. AI-generated fake news. Our analyses are at subject-rating level (i.e., one data point per fake news per subject; $20 \times 988 = 19,760$ observations). We use the Mann-Whitney-U-test for testing our hypothesis. %, $N = $ 19760 observations)
\item For RQ2, we analogously performed hypothesis testing to compare the \emph{WillignessToShare} of human-generated vs. AI-generated fake news. Our analysis is at subject-rating level. We removed \emph{``Don't know''} answers, thereby yielding 17,676 observations. Here, we use the $\chi^2$-test. 

\item For RQ3, we performed a regression analysis to understand which determinants explain why users fall for AI-generated (vs. human-generated) fake news. We constructed two linear mixed-effects regression models, one for each dependent variable, with a subject-level random effect to account for between-subject heterogeneity.\footnote{Our pre-registration originally stated that we account for between-subject heterogeneity through standard errors clustered at the subject level. However, we decided to follow recent research \cite{lutz_affective_2023} and use random effects instead, which introduces a varying-intercept and thus allows for between-subject variation in the tendency to believe in fake news. Both analyses led to qualitatively similar conclusions.} In addition, we control for a number of behavior factors. We control for potential fatigue by including a variable \emph{StatementSequence}, which denotes the current number of already answered statements in the experiment.

The regression models are as follows:
\begin{equation}
\begin{aligned}
\mathit{PerceivedVeracity}_{ij} =\;\; &\beta_0 + u_{0j} + \beta_1 \, \mathit{VaccinationStatus}_{j} + \beta_2 \, \mathit{Concern}_{j}   \\
&+ \beta_3 \, \mathit{NewsConsumption}_{j} + \beta_4 \, \mathit{CRT}_{j}\\
&+ \beta_5 \, \mathit{Age}_{j} +  \beta_6 \, \mathit{Gender}_{j} \\
&+ \beta_7 \, \mathit{PoliticalOrientation}_{j} + \beta_8 \, \mathit{EconomicOrientation}_{j}\\
&+ \beta_9 \, \mathit{EducationYears}_{j}  + \beta_{10} \, \mathit{Religiosity}_{j}  + \beta_{11} \, \mathit{Income}_{j}\\
&+ \beta_{12} \, \mathit{Ethnicity}_{j} + \beta_{13} \, \mathit{StatementSequence}_{ij} + \varepsilon_{ij}, \\
\end{aligned}
\end{equation}
\begin{equation}
\begin{aligned}
\mathit{WillingnessToShare}_{ij} = \;\; &\alpha_0 + u_{0j} + \alpha_1 \, \mathit{VaccinationStatus}_{j} + \alpha_2 \, \mathit{Concern}_{j}   \\
&+ \alpha_3 \, \mathit{NewsConsumption}_{j} + \alpha_4 \, \mathit{CRT}_{j}\\
&+ \alpha_5 \, \mathit{Age}_{j} +  \alpha_6 \, \mathit{Gender}_{j} \\
&+ \alpha_7 \, \mathit{PoliticalOrientation}_{j} + \alpha_8 \, \mathit{EconomicOrientation}_{j}\\
&+ \alpha_9 \, \mathit{EducationYears}_{j}  + \alpha_{10} \, \mathit{Religiosity}_{j}  + \alpha_{11} \, \mathit{Income}_{j}\\
&+ \alpha_{12} \, \mathit{Ethnicity}_{j} + \alpha_{13} \, \mathit{StatementSequence}_{ij} + \varepsilon_{ij}, \\
\end{aligned}
\end{equation} 
where $\mathit{PerceivedVeracity}_{ij}$ and $ \mathit{WillingnessToShare}_{ij}$ are the dependent variables for $i$-th observation of the $j$-th subject, with intercepts $\beta_0$, $\alpha_0$, subject-level random effect $ u_{0j} $, coefficients $\beta_{1}, \ldots$; $\alpha_1, \ldots$, and an error term $\epsilon_{ij}$. 

In our implementation, we use the \emph{lme4} package from R, which uses the restricted maximum likelihood (REML) estimator. We use the REML estimator but also run robustness checks (see Sec 4.5) using logistic regression (for \emph{WillingnessToShare}) and ordinal regression (for \emph{PerceivedVeracity}). As in \cite{pennycook_fighting_2020}, ordinal and numerical variables are $z$-standardized to zero mean and one standard deviation to provide better interpretability. Categorical variables are coded numerically (see Table~\ref{tab:detailQues}). We did not add \emph{SocialOrientation} in the regression as it is highly collinear with \emph{EconomicOrientation}.
\footnote{We repeated the analysis by exchanging the \emph{EconomicOrientation} by \emph{SocialOrientation} which leads to conclusive findings.}
\end{enumerate}

\subsection{Ethical Considerations}

% privacy

We respect the privacy and agency of all people potentially impacted by this work and take specific steps to protect their privacy. To this end, all data were collected in an anonymous form, and any personally identifiable features were removed. All steps during data collection and analyses followed standards for ethical research \cite{rivers_ethical_2014}. All subjects gave informed consent, including for data publication. 

% scope

We further acknowledge that our analysis exposed people to fake news. The specific experimental design was approved by the Institutional Review Board at the University of (anonymized for peer-review). To mitigate potential risks for subjects, we debriefed subjects regarding the aim of the survey and disclosed that all statements were false at the end of the survey and emphasized that they should not rely on them. Moreover, we intentionally chose the COVID-19 pandemic as infection rates have plummeted and the immediate health risk has decreased over time (as opposed to other fake news narratives that currently circulate online). 
% danger RQ

Finally, we acknowledge that our findings could also be used by malicious actors running fake news campaigns. However, there are immediate threats from  AI-generated fake news campaigns for public trust \cite{feuerriegel_research_2023}, and we thus follow earlier calls for research  \cite{feuerriegel_research_2023} that ask for more evidence as a first step to understand risk vectors and design effective policy responses.

%Result

\section{Results}

\subsection{Comparing Perceived Veracity for Human-Generated vs. AI-Generated Fake News (RQ1)}

Our first research question is: ``What is the perceived veracity of human-generated vs. AI-generated fake news?''. Previously, Spitale et al.~\cite{spitale_ai_2023} has conducted an experimental study demonstrating that people find false Twitter/X posts created by GPT-3 more accurate than human-written posts. Inspired by these results, we intend to replicate this investigation in our own experimental setting using fake news (instead of user-generated Twitter/X posts).  

\begin{figure}[htbp]
  \centering
  \includegraphics[width=0.8\linewidth]{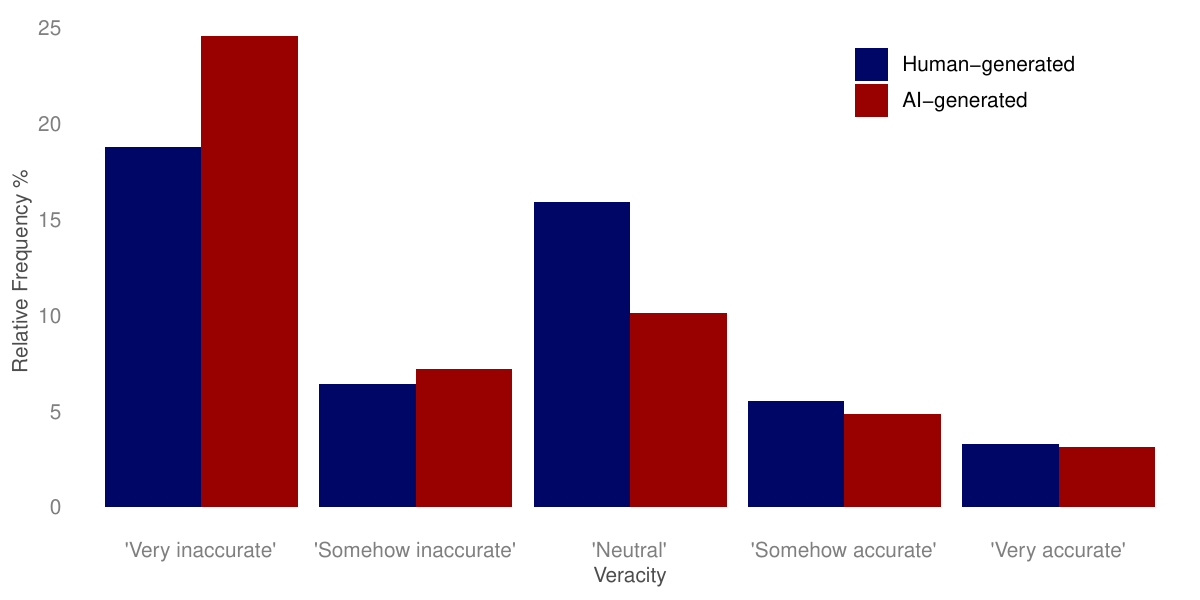}
  \caption{Distribution of perceived veracity scores across human-generated vs. AI-generated fake news.}
  \Description{The figure is a bar chart which shows the relative frequency of chosen perceived veracity options for AI-generated vs. human-generated fake news. For options "very inaccurate" and "somehow inaccurate," the relative frequency of AI-generated is higher, and for the rest, the trend is reversed. }
  \label{fig:veracity_distribution}
\end{figure}

Figure~\ref{fig:veracity_distribution} shows the distribution of subjects' perceived veracity scores grouped by whether the fake news is human-generated or AI-generated. We performed a Mann-Whitney-U-test to examine whether human-generated and AI-generated fake news perceived veracity scores have identical distributions. The result confirmed that the distributions of the perceived veracity scores are not identical across human-generated and AI-generated fake news ($p < 0.001$).
 
The average perceived veracity scores of AI-generated fake news is 19.67\% smaller than that of human-generated fake news. Therefore, subjects perceived human-generated fake news more accurately compared to AI-generated fake news. Consequently, we find a negative answer to our RQ1: AI-generated fake news is considered less accurate than human-generated fake news.

\begin{figure}[htbp]
        \centering
        \includegraphics[width=0.7\linewidth]{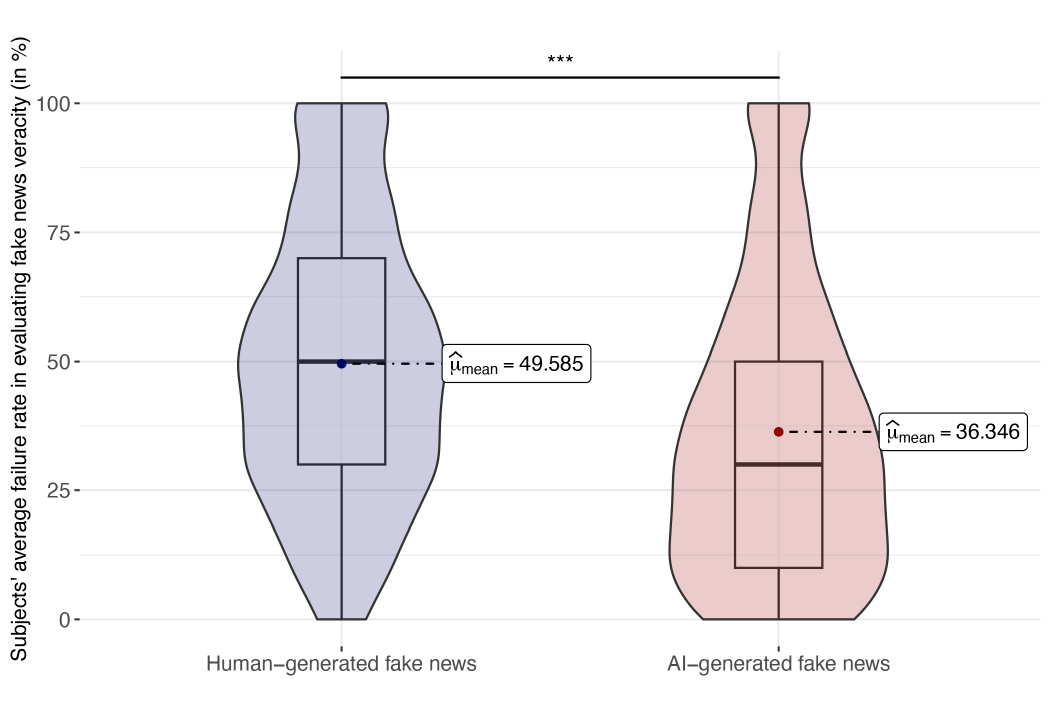}
        \caption{The average rate of incorrect perceived veracity assessments across subjects. Welch's $t$-test shows that human-generated fake news has a significantly ($p < 0.001$) higher rate of incorrect perceived veracity assessments compared to AI-generated fake news. %The effect size for this test $g_{Hedges} = 0.746$.
        }
        \Description{The graph shows the distribution of subjects' failure rate in detecting fake news grouped by fake news generation source. The Welch $t$-test shows the human-generated fake news has significantly ($p < 0.001$) higher rate of incorrect perceived veracity as compared to AI-generated fake news.}
        \label{fig:subject_faliure_rate}

\end{figure}

In addition, we repeat the above analysis using a binary variable of whether users find fake news accurate or not. Formally, the binary variable amounts to 1 if a subject rates fake news as \emph{``somehow accurate''} or \emph{``very accurate''}, and 0 otherwise. Overall, 98.9\% of the users have at least once failed to correctly identify AI-generated fake news as such. We then calculated the relative number of how often users fail to identify fake news as such and thus respond with \emph{``somehow accurate''} and \emph{``very accurate''}. We use Welch's $t$-test to check if there is a significant difference in the average rate of incorrect perceived veracity assessments for human-generated and AI-generated fake news. Figure~\ref{fig:subject_faliure_rate} shows that human-generated fake news has a significantly ($p < 0.001$) higher rate of incorrect perceived veracity assessment. On average, subjects failed 49.6\%  when rating human-generated fake news, while they failed 36.3\% when rating AI-generated fake news. As a robustness check, we also employed a linear mixed-effects regression with subject-level random effects, which corroborated our findings.

Our findings for fake news are contrary to the findings for user-generated social media content from Spitale et al.~\cite{spitale_ai_2023}. In their study, AI could disinform subjects better than humans, implying that the perceived veracity of fake social media posts generated by AI is \emph{higher} than for fake social media posts generated by humans. However, our findings suggest that the perceived veracity of AI-generated fake news is \emph{smaller} than for human-generated fake news. The different findings with regard to the perceived veracity could be rooted in the type of generated content. In their study, Spitale et al.~\cite{spitale_ai_2023} used Twitter/X posts, while we use fake news. Hence, it is likely that user-generated social media content such as Twitter/X posts may be written and generally perceived differently than news. For example, social media posts written by humans may have grammatical errors, which users can use as cues to detect that such posts are fake, while news are generally written without grammatical errors regardless of whether the news originates from humans or an AI).

\subsection{Comparing Willingness to Share for Human-Generated vs. AI-Generated Fake News (RQ2)}

To address RQ2, we present our findings regarding the subjects' willingness to share for fake news. Specifically, we compare the willingness of subject to share human-generated fake news versus their willingness to share AI-generated fake news. This comparison aims to discern any significant differences in the sharing behavior elicited by the source of the fake news. Overall, a substantial portion of the subjects would share human-generated fake news (11.59\%) and AI-generated fake news (11.94\%). We used a $\chi^2$-test to examine whether there is a significant difference between whether subjects share human-generated compared to AI-generated fake news. We found that there is no statistically significant difference between the willingness to share human-generated and AI-generated fake news ($p = 0.69$). Therefore, we find a negative answer to RQ2: there is no difference in willingness to share between human-generated and AI-generated fake news. As a robustness check, we also performed linear mixed-effects regression with subject-level random effects, finding consistent results that there is no statistically significant difference with respect to willingness to share at common significance thresholds.

\begin{figure}[htbp]
  \centering
  \includegraphics[width=0.8\linewidth]{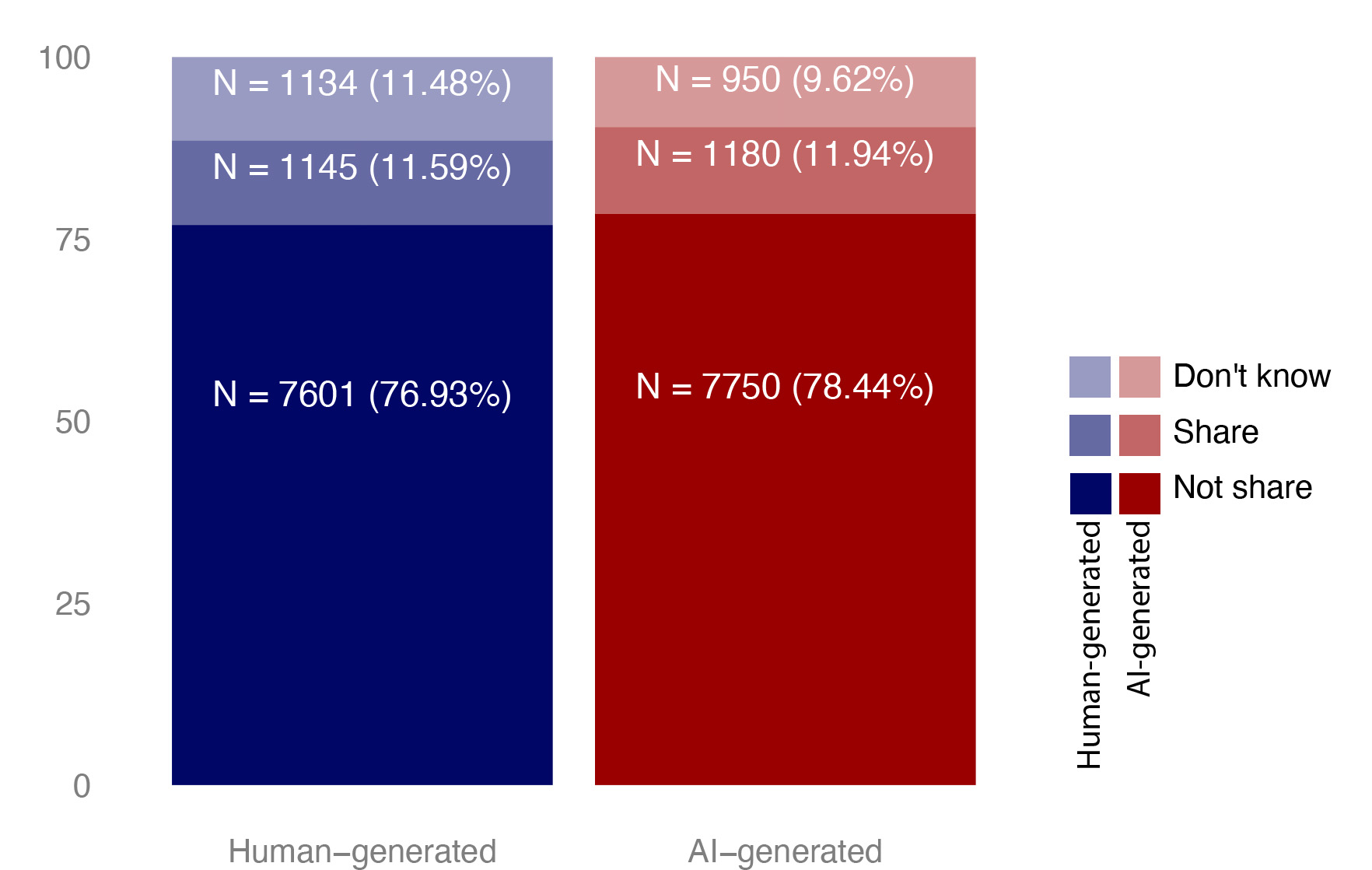}
  \caption{Subjects' fake news willingness to share grouped by source of generation. The $\chi^2$-test shows that there is no significant difference between AI-generated vs. human-generated fake news.}
  \Description{The figure illustrates the result of Chi-squared test on subjects' willingness to share COVID-19 fake news statements.}
  \label{fig:shareStatBar}
\end{figure}

\subsection{Socio-Economic Characteristics as Determinants Explaining Which Users Fall for AI-Generated Fake News (RQ3)}

\begin{figure}[htbp]
  \centering
  \includegraphics[width=0.9\linewidth]{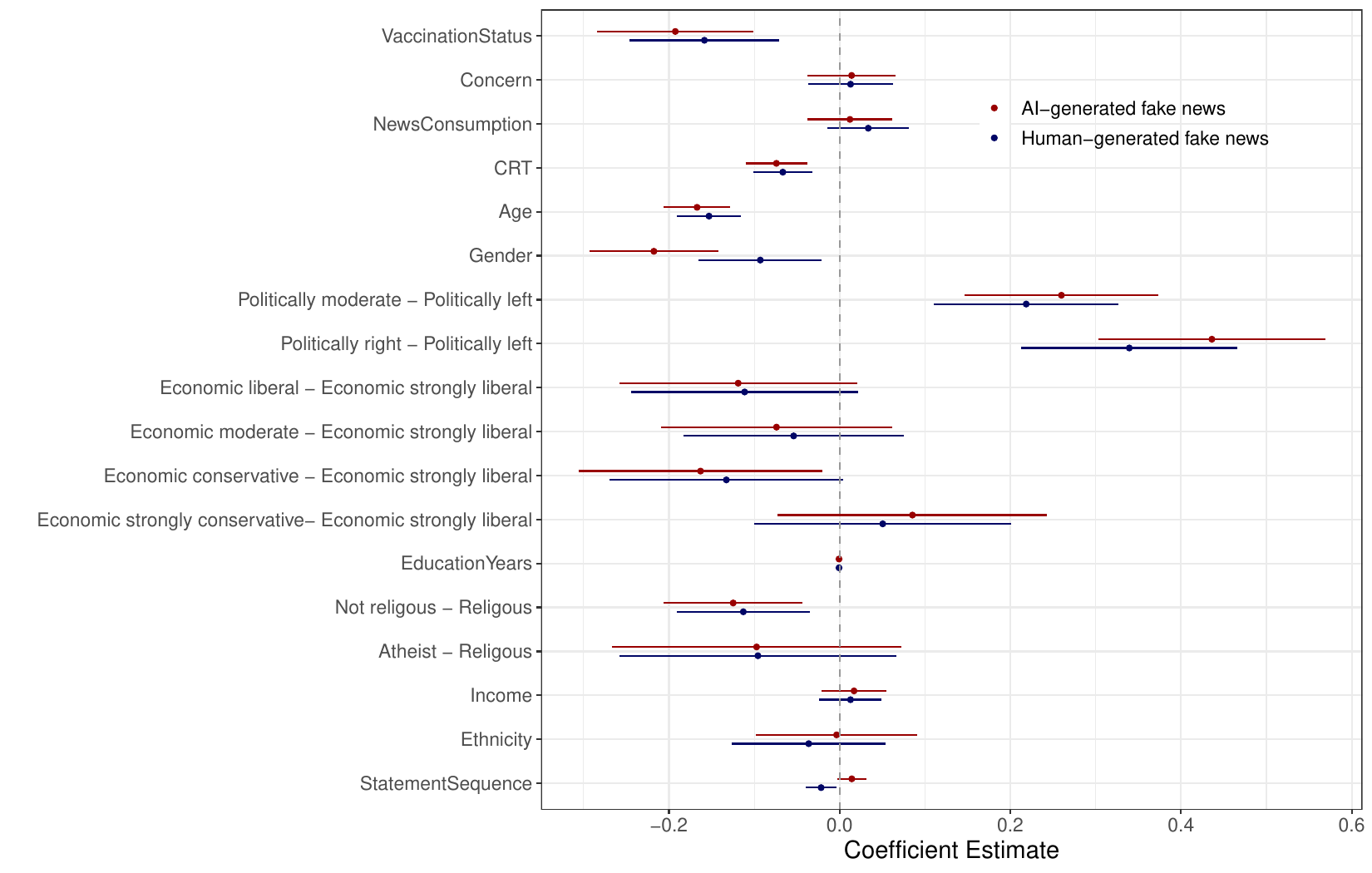}
  \caption{Estimates explaining the perceived veracity through various socio-economic variables using linear mixed-effects regression models. All models include subject-level random effects.}
  \Description{The figure illustrates the graph of perceived veracity regression coefficient estimates.}
  \label{fig:veracity_regression}
\end{figure}

\textbf{Perceived veracity:} We now analyze which socio-economic characteristics are determinants that explain the perceived veracity of users. For this, we use a linear mixed-effects regression with \emph{PerceivedVeracity} as the dependent variable and socio-economic variables as our independent variables. We developed mixed-effects linear regression models with subject-level random effects. We run separate regressions for human-generated fake news and AI-generated fake news, so that we can identify differences in how both are received by subjects. Our findings are shown in Figure~\ref{fig:veracity_regression}.

% rewrite these
\emph{NewsConsumption} and \emph{Concern} are not associated with perceived veracity in any of the models ($p > 0.05$). \emph{CRT} and \emph{VaccinationStatus} are are positive statistically significant and thus are determinants of the perceived veracity, while \emph{CRT} has a negative association with the perceived veracity of the fake news. For a one standard deviation increase in \emph{CRT}, we expect that the perceived veracity reduces by a 7\% standard deviation. Regarding \emph{Religiosity}, we observe that atheists (in reference to religious subjects) do not have any statistical significance in perceived veracity. However, non-religious subjects (in reference to religious subjects) have a negative association with perceived veracity ($p < 0.01$). For \emph{EconomicOrientation}, we observe that the different categories do not have a statistically significant ($p > 0.05$) effect. When it comes to \emph{PoliticalOrientation}, the results show that politically moderate and right subjects perceived the veracity of fake news higher (as compared to politically left). We can see that the association is slightly different for human-generated as compared to AI-generated fake news. Further, \emph{Age} is also statistically significant ($p < 0.001$): a one standard deviation increase in age, we will correspond to a 15.6\% standard deviations decrease in perceived veracity. Regarding \emph{Gender}, females, with reference to males, perceived the veracity of the fake news as lower, and it is also statistically significant ($p < 0.001$). This association exists both for human-generated and AI-generated fake news.

\begin{figure}[htbp]
  \centering
  \includegraphics[width=0.9\linewidth]{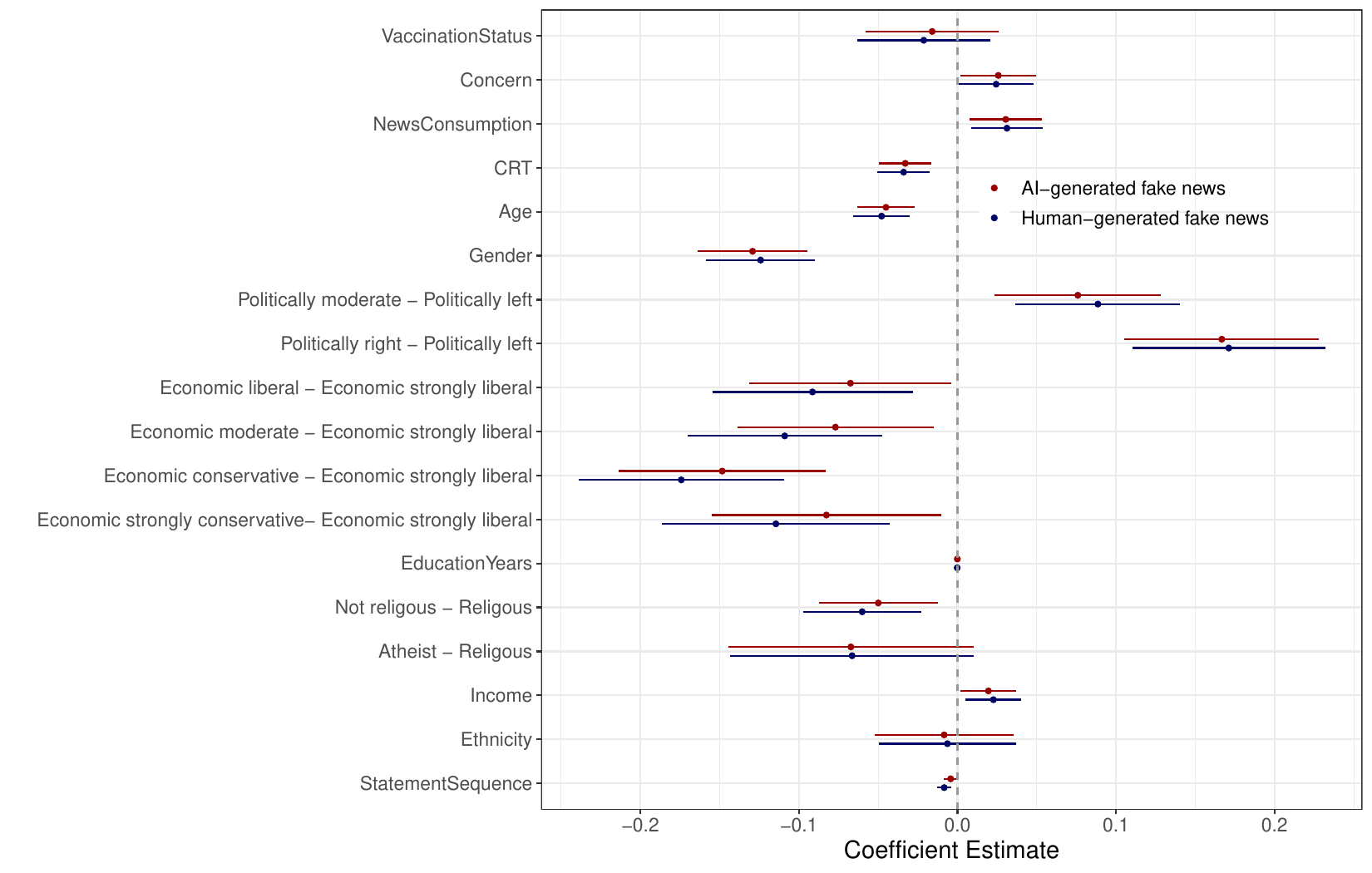}
  \caption{Estimates explaining the willingness to share through various socio-economic variables using linear mixed-effects regression models. All models include subject-level random effects.}
  \Description{The figure illustrates the graph of willingness to share regression coefficient estimates.}
  \label{fig:share_reg}
\end{figure}

\textbf{Willingness to share:} We now analyze which socio-economic variables are associated with \emph{WillingnessToShare} (see Figure~\ref{fig:share_reg}). In contrast to \emph{PerceivedVeracity}, our findings show that \emph{NewsConsumption} ($p < 0.001$) and \emph{Income} ($p < 0.01$) are significantly positively associated with \emph{WillingnessToShare}. However, the effect size is small. Interestingly, a larger values for \emph{Consumption} and \emph{Concern} correspond to a larger \emph{WillingnessToShare}. Furthermore, \emph{CRT}, \emph{PoliticalOrientation},  \emph{Religiosity}, \emph{EducationYears}, \emph{Ethnicity}, and \emph{Age} follow the same pattern as for \emph{PerceivedVeracity}. Although \emph{Gender} is still statistically significant ($p < 0.001$), the estimate for human-generated fake news and AI-generated fake news is very close, unlike as for \emph{PerceivedVeracity} as the dependent variable.

\subsection{Extensions}

\begin{figure}[htbp]
        \centering
        \includegraphics[width=0.8\linewidth]{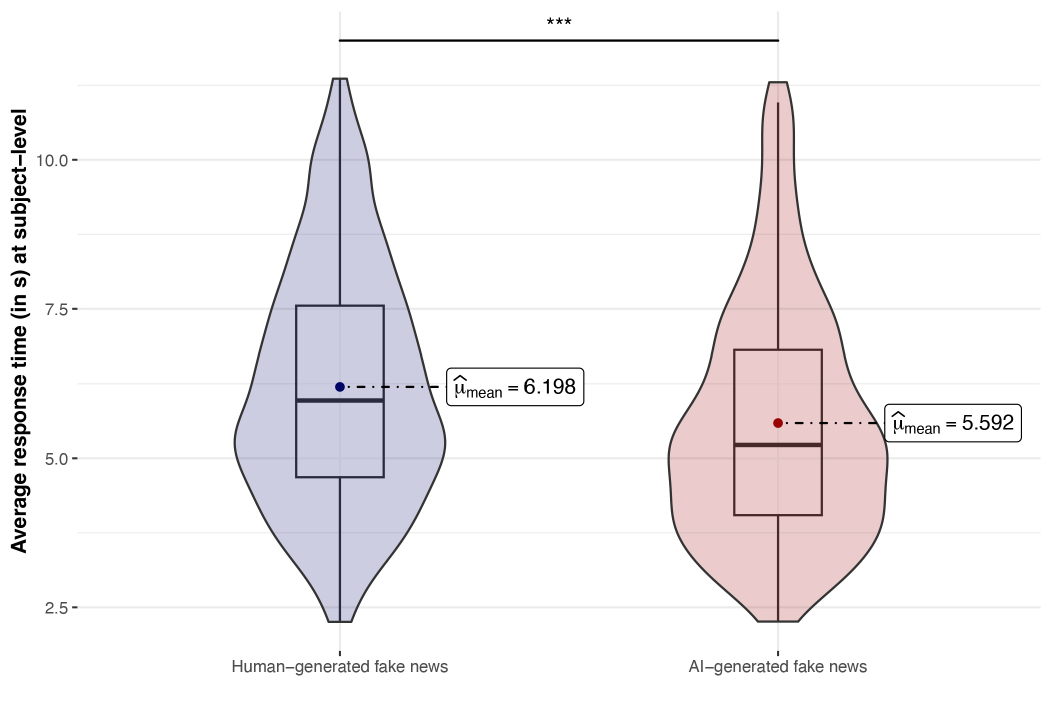}  
        \caption{Subjects' average response time in seconds grouped by human-generated vs. AI-generated fake news. The upper bound of the data is capped at 90th percentile to avoid the influence of outliers. A Welch's $t$-test shows a significant difference in the means of human-generated vs. AI-generated fakes news ($p \simeq$ 0). However, as the plot shows, the effect size is fairly small.}
        \Description{The graph shows the distribution of subjects' response time in seconds grouped by fake news generation source. The Welch t-test shows a significant difference in the means of human-generated vs. AI-generated fakes news ($p \simeq$ 0). However, as the  \emph{Hedges' g} is equal to 0.302 we  concluded that the effect size is small.}
    \label{fig:response_time_dist}
\end{figure}

Users may vary in their attention. To control for that, we
consider the subjects' average response time as a potential determinant in how accurately users can assess the veracity of fake news. To this end, we calculated the average response time at the subject-level for human-generated and AI-generated fake news. To reduce the influence of outliers on our estimate, we capped the upper bound of the average response time to 90th percentile. Figure~\ref{fig:response_time_dist} shows the average response time for each subject for both human-generated and AI-generated fake news statements. A Welch $t$-test confirms that subjects responded to AI-generated fake news significantly faster than human-generated ($p < 0.001$). However, the difference in the means is comparatively small ($0.604$ seconds).\footnote{We also repeated the above regression analysis using the response time as an addition control for between-subject heterogeneity in attention but we arrived at conclusive findings.}

\subsection{Robustness Checks}

We followed our pre-registration and conducted several robustness checks to validate our results. (1)~We calculated the variance inflation factor (VIF) for all regression models. All VIFs are below the critical threshold of five \cite{akinwande_variance_2015}, implying that there is no concern due to multicollinearity. (2)~One of our dependent variables is ordinal (i.e., {\emph{PerceivedVeracity}}), and the other is binary (i.e., {\emph{WillingnessToShare}}). We nevertheless chose ordinary least squares for reasons of better interpretability. We additionally repeated our analysis with a logistic regression and an ordinal regression, respectively, yielding similar findings. (3)~We excluded subjects who either failed the second set of attention tests or indicated that they selected answers randomly or searched the Internet during the experiment. When including the subjects instead, we arrived at consistent findings. In sum, all of the robustness checks supported our findings.

%Discussion

\section{Discussion}

\subsection{Relevance}

% threats

The phenomenon of AI-generated fake news has emerged as a critical concern. While the concept of fabricating information, such as deepfakes \cite{burkhardt_combating_2017}, is not new, the advent of large language models and other generative AI techniques make it easy to create fake news at an unprecedented scale \cite{feuerriegel_research_2023,menz_health_2024}. In particular, AI-generated fake news could make it difficult for online users to discern what is true and what is false, thus amplifying the threat vectors from fake news with detrimental consequences for public trust. Hence, research such as ours is needed to explore the vulnerabilities of individuals to AI-generated fake news. 

% new "cues"

Large language models possess the capability to produce fake news that is difficult for humans to detect. In particular, we have demonstrated empirically that 98.9\% of the subjects have at least once fallen for the AI-generated fake news in our sample. This finding can be attributed to the ability of state-of-the-art large language models to mimic the style and content of reputable sources. As such, many cues that humans have previously used to detect fake content are no longer valid. For example, checking the grammatical accuracy within texts is no longer sufficient since large language models can write virtually error-free textual content.

% novelty

Previous research has already analyzed whether humans can discern AI-generated content from human-generated content, such as for dating profiles, hostility profiles, or letters \cite{jakesch_human_2023,kreps_all_2022}). Here, humans are often unable to accurately detect AI-generated content. However, the aforementioned results are based on regular, truthful content, whereas the phenomenon of fake news introduces additional complexity in that a false statement around which to write fake news must also be fabricated. The latter presents the focus of our study. Previously, the \emph{perception} of users in terms of perceived veracity of human-generated vs. AI-generated fake news was analyze \cite{spitale_ai_2023}, while our novelty is to analyze the \emph{behavior} in terms of willingness to share. The latter is crucial to understand, as sharing is an important precondition for fake news to go viral and thus reach a large audience. Importantly, as we show in our study, perceived veracity and sharing behavior are only weakly correlated and thus capture different activities of online behavior.

\subsection{Implications of Findings}

Our paper has three main findings. First, even though our results highlight that AI-generated fake news is perceived to be less accurate than human-generated fake news, a large percentage of subjects fall for AI-generated fake news. Second, we do not find a statistically significant difference in the willingness of subjects to share human-generated vs. AI-generated fake news despite the relatively large sample size, highlighting the risk that fabricated content may go easily viral. Third, we find that (some) socio-economic factors explain whether users fall for AI-generated fake such as age and political orientation. In our setting with fake news from the COVID-19 pandemic, subjects who disclose to be conservatives were especially prone to fall for AI-generated fake news. This finding indicates that people from different socio-economic backgrounds could fall for AI-generated fake news also in different ways.

As implications, we see three potential directions that are valuable to counter risks due to AI-generated fake news. (1)~Novel cues are needed that users can leverage to discern true from fake content, yet such cues must not be easily gamed by AI technologies. Specific examples are digital watermarks or community-based fact-checks (e.g., Community Notes at Twitter/X). (2)~Our findings highlight the importance of making people aware of the risks of AI-generated fake news, implying that tailored media literacy trainings are needed. For example, one could train online users in how generative AI works so that they check the perceived veracity of online content more carefully. (3)~Online users may fall for AI-generated fake news without noticing, because of which regulatory approaches may be needed \cite{goldstein_generative_2023}. Notwithstanding, as our regression results show, the vulnerabilities to AI-generated fake news can also be pronounced for certain marginalized groups (e.g., with a lower cognitive reflection test score, young age).  

\subsection{Limitations}

As with other works, our work has several limitations that open up avenues for future research. First, our study is based on fake news about COVID-19. While the COVID-19 pandemic had large negative impacts on public health, the results may be different for other contexts (e.g., armed conflicts) and other forms of disinformation (e.g., propaganda, rumors, hoaxes). Hence, future research may replicate our analyses in such varied contexts. Second, we only measure the willingness to share and not actual decisions to share. However, this is consistent with prior research \cite{pennycook_fighting_2020}, and, on top of that, it would be highly unethical to ask subjects to actually share fake news in their social media account. Third, our results are based on GPT-4, a large language model with state-of-the-art performance \cite{openai_gpt-4_2023}. However, the research landscape around generative AI technologies is quickly evolving, and new models with better capabilities will be developed over time (including generative AI technologies for multi-modal outputs). Here, it will be interesting to repeat our research with new models and other modalities such as fake images and videos.

\subsection{Conclusion}

Generative AI can give rise to harmful applications when it is used to generate fake news. Yet, it is unclear how online users interact with AI-generated fake news. We thus provided a large-scale analysis with experimental evidence, finding that humans are likely to share AI-generated fake news but that the sharing propensity is similar for human-generated and AI-generated fake news.

% Convert yash refrences name after importing to zotero
\bibliographystyle{ACM-Reference-Format}
\bibliography{references.bib}
\end{document}

\pagestyle{fancy}
\lhead{Fake news corpus}

\begin{sidewaystable}
\caption{Full list of fake news in our corpus.} 
\label{tab:example_fakenews} 
\renewcommand{\arraystretch}{1.3}
\footnotesize
\begin{tabular}{lll} 
\toprule
Content & Label & Source\\ 
\midrule 
\emph{``The coronavirus has been found in the one place people never expected, toilet paper.''} & Human-generated & Snopes\\
\emph{``At the World Economic Forum, Moderna CEO admitted that his company produced 100,000} \\ \emph{ doses of COVID-19 vaccines in 2019, before the pandemic started.''}
& Human-generated & Snopes\\
\emph{``COVID-19 vaccines threaten people's fertility, no matter their type of reproductive systems.''} & Human-generated & Snopes\\
\emph{``COVID-19 mRNA vaccines contain graphene oxide.''} & Human-generated & Snopes\\
\emph{``If you have natural immunity, taking the vaccine is more likely to harm you severely.''} & Human-generated & PolitiFact\\
\emph{``The second booster has 8 strains of HIV.''} & Human-generated & PolitiFact \\
\emph{``Vatican confirms Pope Francis and two aides test positive for coronavirus.''} & Human-generated & Pennycook et.al$^{*}$\\
\emph{``Experts think bats are the source of the Wuhan Coronavirus. At least 4 pandemics have originated} \\ \emph{ in these animals.''} & Human-generated & Pennycook et.al$^{*}$\\
\emph{``Federal Emergency Management Agency proposes martial law to contain virus.''} & Human-generated & Pennycook et.al$^{*}$ \\
\emph{``Coconut oil's history in destroying viruses, including coronaviruses.''} & Human-generated & Pennycook et.al$^{*}$ \\
\midrule
\emph{``COVID-19 can be treated by taking hydroxychloroquine and ivermectin.''}  & AI-generated & GPT-4\\ 
\emph{``The coronavirus was created in a lab as a bioweapon.''} & AI-generated & GPT-4\\
\emph{``The coronavirus can be prevented by eating garlic or taking vitamin C supplements.''} & AI-generated & GPT-4 \\
\emph{``The coronavirus only spreads in cold weather or places with low humidity.''} & AI-generated & GPT-4 \\
\emph{``The coronavirus can be detected by holding your breath for 10 seconds or by smelling} \\ \emph{ something strong.''} & AI-generated & GPT-4 \\
\emph{``COVID-19 vaccine causes infertility in women and men.''} & AI-generated & GPT-4\\
\emph{``COVID-19 can be transmitted by mosquitoes and other insects.''} & AI-generated & GPT-4\\
\emph{``COVID-19 is less deadly than the flu and does not require any special measures.''} & AI-generated & GPT-4\\
\emph{``COVID-19 vaccines contain microchips that can track your location and control your mind.''} & AI-generated & GPT-4 \\
\emph{``COVID-19 can be spread by pets and animals to humans.''} & AI-generated & GPT-4 \\
\bottomrule
Note: * [Pennycook et al.(2020)] Gordon Pennycook, Jonathon McPhetres, Yunhao Zhang, \\ Jackson G. Lu, and David G. Rand. 2020. Fighting COVID-19 Misinformation on Social Media: \\Experimental Evidence for a Scalable Accuracy-Nudge Intervention. Psychological Science 31, 7 \\(July 2020), 770– 780. https://doi.org/10.1177/0956797620939054 Publisher: SAGE Publi- cations Inc.
\end{tabular}
\end{sidewaystable}